# Lithium depletion in solar analogs: age and mass effects

Anne Rathsam ⬤,⋆ Jorge Meléndez and Gabriela Carvalho Silva

*Instituto de Astronomia, Geofísica e Ciências Atmosféricas, Universidade de São Paulo, Rua do Matão 1226, São Paulo 05508-090, Brazil*

## ABSTRACT

The main goal of this work is to evaluate the correlation between Li abundance, age, and mass. Using high-quality ESO/HARPS data (R ≃ 115 000; 270 ≤ SNR ≤ 1000), we measured Li abundances via spectral synthesis of the 6707.8 Å $^7$Li line in 74 solar twins and analogs. Our joint analysis of 151 Sun-like stars (72 from our sample plus 79 solar twins from a previous study) confirms the strong Li abundance–age correlation reported by other works. Mass and convective envelope size also seem to be connected with Li abundance but with lower significance. We have found a link between the presence of planets and low Li abundances in a sample of 192 stars with a high significance. Our results agree qualitatively with non-standard models, and indicate that several extra transport mechanisms must be taken into account to explain the behaviour of Li abundance for stars with different masses and ages.

**Key words:** techniques: spectroscopic – stars: abundances – stars: evolution – stars: low-mass – planetary systems – stars: solar-type.

## 1 INTRODUCTION

Lithium is an extremely important element for stellar astrophysics because it gets depleted at somewhat low temperatures (~2.5 × 10$^6$ K) which are easily reached in the interior of stars. The base of the convective zones of main sequence solar-type stars have a temperature of around 2 × 10$^6$ K and are thus not able to burn Li. Therefore, the Li abundance of a single solar-type star should not change as the star evolves along the main sequence according to the standard solar model, since it does not consider any transport mechanism besides convection.

However, old field solar analogs present a lower Li abundance (e.g. Carlos, Nissen & Meléndez 2016; Carlos et al. 2019), indicating that part of the Li present in the stars is taken towards inner (and hotter) regions through non-standard mixture processes. Another evidence of the dependence between Li depletion and age is that the Sun's photospheric value of Li abundance is around 150 times lower than the meteoritic value (adopted as the value of Li abundance during the formation of the Solar system; Asplund et al. 2009), and zero-age main sequence (ZAMS) clusters like the Pleiades show that although there is some pre-main-sequence depletion (e.g. Bouvier et al. 2018), it cannot explain the low solar Li abundance alone.

Many additional transport mechanisms were invoked in an attempt to explain the observed Li destruction, such as convective overshooting (Xiong & Deng 2009), in which the material reaches the convective envelope (that acts as a 'break') with a certain momentum and is carried towards inner regions due to its inertia; convective settling (Andrássy & Spruit 2015), in which the material reaches the Li burning zone due to the low entropy of the downflow; atomic diffusion (do Nascimento et al. 2009), in which the atoms are taken deep inside the star via gravitational settling; rotation-induced mixing

(Charbonnel, Vauclair & Zahn 1992; Pinsonneault 1994; Deliyannis 2000; do Nascimento et al. 2009; Denissenkov 2010); and mixture due to gravity waves in the stellar interior (Charbonnel & Talon 2005), that redistribute the angular momentum effectively along the star.

Non-standard models that take into consideration one or more extra mechanisms of transport of material are able to predict a reduction of the Li abundance with age. Nevertheless, no model has been able to simultaneously explain every observed characteristic regarding Li abundance in different stellar types and metallicities, such as the Spite plateau (Spite & Spite 1982), that shows that old stars belonging to the Galactic halo reach a maximum value for Li abundance that is below what is predicted by primordial nucleosynthesis – although the Li destruction is not expected in this type of star due to their shallow convective zones – and the Li dip (first observed in the Hyades cluster by Boesgaard & Tripicco 1986), that shows that stars with effective temperatures between 6200 and 7000 K in open clusters present an abrupt decay of their Li abundances in comparison with stars warmer or colder than this interval.

It has been suggested that Li is depleted mostly during the pre-main-sequence phase, when the stellar convective envelope is deeper and thus the overshooting effects are stronger (Thevenin et al. 2017); yet, this disagrees with the observed smooth decay of Li abundance with stellar age (Monroe et al. 2013; Carlos et al. 2019). This indicates that the extra mixing process does not reach too deep inside the star, because the nuclear fusion grows exponentially with temperature, and the temperature gets higher as we approach the stellar core. Additionally, Tucci-Maia et al. (2015) found roughly constant beryllium abundances in field solar twins, which also points towards a shallow extra mixing mechanism. Similar to Li, beryllium is depleted at a relatively low temperature (~3.5 × 10$^6$ K), although higher than the temperature needed to destroy Li. Therefore, the additional transport method reaches deep enough to burn Li in solar twin stars, but not so deep as to begin beryllium depletion.

⋆ E-mail: annerathsam@usp.br

Another possibility that has been explored is that Li depletion is related to the formation and presence of planets (Israelian et al. 2009; Delgado Mena et al. 2014; Figueira et al. 2014; Gonzalez 2015), which is supported by the finding that the Sun is more Li-depleted than solar twins of similar age (Carlos et al. 2019). This idea, however, has been contested by other works, such as Baumann et al. (2010), Ghezzi et al. (2010), and Bensby & Lind (2018). The presence of planets is suggested to also cause an increase in the stellar Li abundance in case of planet engulfment events (Sandquist et al. 2002; Meléndez et al. 2017; Spina et al. 2021). At the moment, more observations are required in order to determine the correlation between Li abundance and the presence of planets.

Besides age, there are other parameters that may affect Li abundance, such as stellar mass and metallicity, which is related to the size of the convective envelope, because both metal rich and low-mass stars present deeper convection zones. The correlation between Li abundance and metallicity for solar analogs is explored by Martos et al. (2023), where they report that stars with higher metal content are able to destroy Li more effectively. However, the dependence of stellar mass (sometimes adopting the effective temperature as a proxy) on Li abundance has been less explored in the literature (examples include Xiong & Deng 2009; Carlos et al. 2016; Carlos et al. 2019; Borisov, Prantzos & Charbonnel 2022, or different works on open clusters, such as Soderblom et al. 1993 for the Praesepe cluster, Pace et al. 2012 for M67, Zappala 1972; Sestito & Randich 2005; Castro et al. 2016; Dumont et al. 2021).

Another intriguing aspect of the relationship between Li abundance and age is that open clusters, that have the advantage of very precise age determinations, do not seem to follow the same behaviour as field stars. Works such as Randich, Sestito & Pallavicini (2003) for NGC 188 (6–8 Gyr old) and Sestito, Randich & Pallavicini (2004) for NGC 752 (∼2 Gyr old) showed that Li depletion in open clusters seems to halt at some point around 2 Gyr. However, the studied samples were composed mostly of stars slightly more massive than the Sun (see e.g. fig. 3 in Randich et al. 2003 and fig. 2 in Sestito et al. 2004), which are stars with thinner convective zones, that will naturally deplete less Li. In contrast, the work of Carlos et al. (2020), that analysed solar twins from the solar age and solar [Fe/H] cluster M67, concluded that the solar twins from M67 follow the Li-age relationship found for field solar twins in Carlos et al. (2019). Carlos et al. (2020) also reported a large scatter in Li abundances for M67, compatible with what is seen in field stars, which could be the result of small differences in mass, metallicity, or rotational velocity. Nevertheless, this scatter was not observed in NGC 188 nor NGC 752. More observations of Li in open cluster stars of (sub)solar masses is necessary to solve this apparent difference in behaviour, which will be the subject of a future study.

In this work, we intend to evaluate the combined effect of stellar age and mass on the Li abundance in a sample of 153 field solar-type stars (74 new stars plus the sample of Carlos et al. 2019). We hope to encourage the development and improvement of non-standard stellar interior models that will be able to reproduce the observed behaviour.

## 2 METHODS

### 2.1 Sample selection

The sample was composed of 74 field solar twins and analogs belonging to the Galactic disc (mainly the thin disc), selected from an updated catalogue of stellar parameters from Ramírez & Meléndez (2005). To include only stars with metallicities close to solar, we restricted our sample to the metallicity range of $-0.15$ dex ≤ [Fe/H]

$\leq +0.15$ dex. We also excluded stars with a $V$ magnitude above 10, as they are too faint and difficult to observe with high-precision. Then, we selected solar twins and solar analogs with masses below 1 $M_\odot$, as an attempt to increase and homogenize the sample of solar twins of Carlos et al. (2019), which was included in the analysis.

We then excluded stars classified as spectroscopic binaries in the SIMBAD data base,[1] as their spectra could be contaminated by their companion, complicating the analysis. We also verified the RUWE (renormalized unit weight error) from the third Gaia data release (DR3), and found that only two stars from our sample had RUWE > 1.4, which can be an indication of multiplicity (HD 169822, RUWE = 18.5, and HD 16541, RUWE = 3.3). These two stars were analysed, but were excluded from the plots and fits presented in this work.

#### 2.1.1 Solar twins

The sample of Carlos et al. (2019) lacked solar twins younger than the Sun with $1.00 \leq M/M_\odot \leq 1.02$. Therefore, we selected stars in this mass range that had log $g \geq 4.50$ dex, which restricted the selection to stars much younger than the Sun (since log $g_\odot = 4.44$ dex).

#### 2.1.2 Low-mass stars

Since lower-mass stars present deeper convective zones, we expect this type of stars to have undergone more Li depletion in comparison with similar-age, higher-mass stars. Thus, they can give us important insights in the study of stellar interior and mixing processes. Therefore, we performed a selection of stars with the described restriction in [Fe/H] and V and with masses ranging from 0.85 to 1.00 $M_\odot$, to increase the coverage in mass of the sample of solar twins of Carlos et al. (2019).

### 2.2 Data processing

The spectra for all our stars were taken from ESO's (European Southern Observatory) public data base,[2] observed with the HARPS (High Accuracy Radial Velocity Planet Searcher) spectrograph. The initial query was for individual SNRs (signal-to-noise ratios) ranging between 20 (to avoid noisy individual spectra) and 370 (to ensure they would not be saturated). We also obtained HARPS spectra for the Sun, reflected in the surfaces of the Moon and the asteroid Vesta.

The Doppler correction of the spectra was performed using the radial velocity value given by HARPS' automatic reduction files and bash commands. Then, the spectra of each star were combined to reduce the contamination from telluric lines (which works especially if the observations are from different dates) and to reach a final SNR higher than the SNR of individual spectra. This was performed using IRAF's[3] SCOMBINE task, adopting the median counts as weights. In this way, we were able to achieve combined spectra with SNRs between 270 and ∼1000, allowing us to perform our measurements with high precision.

Before the normalization, the combined spectra were split into seven parts using IRAF's SCOPY command. Each part was normalized with the CONTINUUM command, adjusting a cubic spline polynomial. The order of the polynomial was similar for a given segment, and was determined iteratively, evaluating not only the residuals but also the visual fit to the continuum. In the end, the normalized parts were recombined with the SCOMBINE command.

---

[1]SIMBAD Astronomical data base, http://simbad.cds.unistra.fr/simbad/.
[2]ESO Science Archive Facility, http://archive.eso.org/.
[3]Image Reduction and Analysis Facility, http://iraf.nao.ac.jp/.

## 2.3 Atmospheric parameters

The atmospheric parameters ($T_{eff}$, log $g$, [Fe/H], and microturbulence velocity $v_t$) were determined via the line-by-line differential spectroscopic method (Bedell et al. 2014; Meléndez et al. 2014), adopting the Sun as the reference star. We measured equivalent widths of iron (Fe I and Fe II) lines individually through IRAF's task SPLOT for all stars in our sample and the Sun (using mostly Vesta's spectra, that usually presented the least telluric contamination, but adopting the Moon's spectra when needed). The measurements were performed adjusting a gaussian curve to the line profile and deblending lines when necessary. To determine the continuum for each spectral line, we overplotted the spectra for a few representative stars in our sample with the spectrum of the Sun in a window of 5 Å centred in the line of interest. Our line list was selected from Meléndez et al. (2014).

With the measured Fe lines, the atmospheric parameters were found via the PYTHON code q² (*qoyllur-quipu*,[4] Ramírez et al. 2014). The parameters are estimated iteratively, and the final result is achieved when the excitation and the ionization equilibria are simultaneously reached. For calculations, q² uses the stellar line analysis code MOOG (Sneden 1973; ABFIND routine) and the Kurucz model atmospheres ATLAS9 (Castelli & Kurucz 2003).

## 2.4 Ages and masses

The ages and masses of the stars in our sample were determined with the q² code, using the stellar atmospheric parameters previously found spectroscopically and Yonsei–Yale isochrones (Yi et al. 2001; Kim et al. 2002). Through the comparison between the 'observed' spectroscopic parameters and the 'theoretical' parameters from isochrones, q² produces a probability distribution for the ages and masses. In this work, we used the most probable value for the age and mass.[5] The uncertainties were adopted as the 1-$\sigma$ range around the most probable value (or the standard deviation for HD 201219).

To find the masses and ages, we applied a correction to account for the contribution of alpha elements abundance to the global metallicity of the star, following the method of Spina et al. (2018). In their work, they added to the [Fe/H] value an offset of $\Delta$[Fe/H] $= \log_{10}(0.64 \times 10^{[\alpha/Fe]} + 0.36)$ (Salaris, Chieffi & Straniero 1993), with an additional offset of $-0.04$ dex so that for a star with solar parameters we obtain both the solar age and the solar mass.

As a proxy for the alpha element abundance, we used the magnesium (Mg) abundance, and measured equivalent widths of five Mg lines (from Meléndez et al. 2014's line list) with IRAF, following the same procedure adopted for the Fe lines, taking special care with potential telluric blends for the Mg lines around ~6319 Å. The Mg abundance was calculated through q², also adopting the ATLAS9 atmospheres.

Then, the metallicity used for the calculation of masses and ages was [M/H], determined by

$$[M/H] = [Fe/H] - 0.04 + \log_{10}(0.64 \times 10^{[Mg/Fe]} + 0.36). \quad (1)$$

With this approach, we must apply a last correction in the metallicity for stars with $-1 \leq$ [Fe/H] $\leq 0$. This is because the isochrones in this range of metallicity were adjusted to follow the trend of Galactic chemical evolution, using the equation from Salaris et al. (1993) with [$\alpha$/Fe] $= -0.3 \times$ [Fe/H] (Meléndez et al. 2010).[6] Therefore, we subtracted $\log_{10}(0.64 \times 10^{-0.3 \times [Fe/H]} + 0.36)$ from the [M/H] calculated via equation (1) for these stars.

For consistency, the ages and masses for the solar twins from Spina et al. (2018) and Carlos et al. (2019) were recalculated in the same manner, and are reported in Martos et al. (2023).

The method we adopted for determining masses and ages with q² uses parallax and $V$ magnitudes, as well as the spectroscopic parameters. We used the Gaia DR3 parallaxes and the $V$ magnitudes from SIMBAD. In the cases of stars without $V$ magnitudes available, we used Gaia's DR3 $G$ magnitude and the (BP–RP) colour index transformed via

$$V = (1.0008 \pm 0.0015) \times G + (0.27 \pm 0.03) \times (BP-RP)$$
$$+ (-0.020 \pm 0.008) \times [Fe/H] + (-0.07 \pm 0.03), \quad (2)$$

which is based on a fit performed with stars with 5149 K $\leq T_{eff} \leq$ 6210 K and $-0.71$ dex $\leq$ [Fe/H] $\leq 0.34$ dex (RMS = 0.016). For stars with $2 < G < 8$, the $G$ magnitude was first corrected following the recommendation of Riello et al. (2021; their equation C.1).

We found typical (median) errors in masses and ages of about 0.01 $M_\odot$ and 0.4 Gyr. This was possible due to the precise Gaia DR3 parallaxes and the differential spectroscopic method adopted, that is associated with very low errors in the atmospheric parameters.

For our sample, we found that it was not necessary to take into account the extinction to derive masses and ages, as the stars fall within the Local Bubble, a local low-density irregular cavity, with closest extinction located at about 80 pc from the Sun (Cox & Reynolds 1987; Welsh et al. 2010). Through inversion of parallaxes, we found that all the stars in our sample are within 60 pc from the Sun, with a median distance of 33 pc. Therefore, since the stars are within the Local Bubble, the effect of extinction can be safely neglected, without introducing any systematic bias in our results.

The mass of the convective envelope was estimated via interpolation using the TERRA code (Galarza, Meléndez & Cohen 2016), which is based on theoretical values from Spada et al. (2017) and considers the effect of the stellar mass and metallicity on the convective mass.

## 2.5 Li abundances

The Li abundances[7] were determined through the spectral synthesis of the 6707.8 Å $^7$Li line with the 2019 version of the 1D LTE MOOG code. The input model atmospheres were interpolated from the Kurucz model atmospheres ATLAS9. The macroturbulence velocity and the projected rotational velocity ($v \sin i$) were calculated using the relations introduced by dos Santos et al. (2016). These relations require the full width at half maximum (FWHM) from the HARPS' CCF (cross-correlation function); we adopted the average of the FWHM values given for each observation (of the same star).

For the spectral synthesis, we adjusted iteratively the abundance of Li and nearby atomic and molecular species. The final synthetic spectrum (with its corresponding Li abundance) is achieved when the deviation from the observed spectrum is minimized. The line list adopted was taken from Meléndez et al. (2012), which takes into account blends and the hyperfine structure of the Li line. We did not consider the contribution of $^6$Li in the spectral synthesis, as it has a much lower abundance in the Sun (Asplund et al. 2009). Examples of the fit to the observed spectrum are shown in Fig. 1.

---

[5]For HD 201219 (marked with $*$ in Table A1), a most-probable age was not well-defined, and therefore we used the mean age of the solutions found.

[6]The isochrones suffered a different correction in the case of [Fe/H] $< -1$, but since all our stars have larger metallicity values, this was not relevant for us.

[7]In this work, we will consider A(Li), defined as A(Li) =log ($N_{Li}/N_H$) + 12.

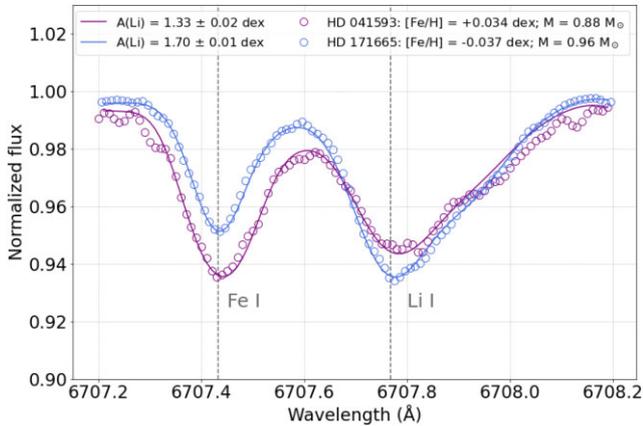

**Figure 1.** Example of the spectra of two stars with similar age but different mass, HD 041593 ($4.5^{+0.9}_{-2.2}$ Gyr) and HD 171665 ($4.8^{+0.4}_{-0.2}$ Gyr). Open circles are the observed spectrum and the solid line is the synthetic spectrum.

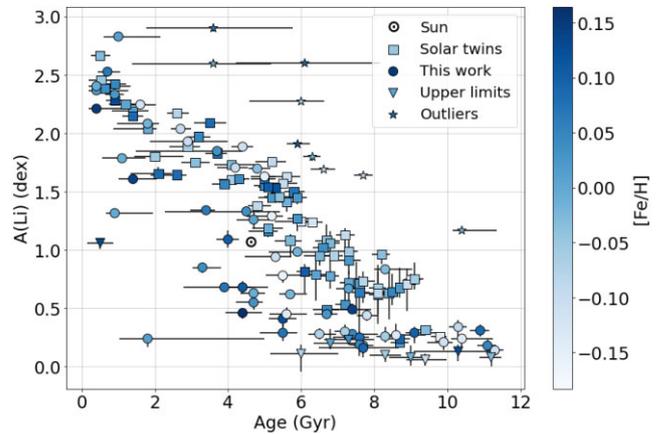

**Figure 2.** A(Li) versus age for the stars in our sample and the solar twins from Carlos et al. (2019); colour-coded by [Fe/H].

To estimate the Li abundance errors, we considered the observational error (related to the quality of the data and depth of the absorption feature) and the systematic error (related to the uncertainties of the atmospheric parameters). The observational errors were determined by finding the deviation of the minimum and maximum fits from the best fit. The systematic errors were found by re-fitting the spectrum with a model atmosphere developed by adding/subtracting the uncertainty of each of the atmospheric parameters at a time while maintaining the others fixed. These errors are added in quadrature to obtain the systematic error. The total error is then $\sigma_{A(Li)} = \sqrt{\sigma_{obs}^2 + \sigma_{sys}^2}$. The typical (median) Li abundance error is 0.04 dex. This error is dominated by the observational error, due to the low uncertainties of the atmospheric parameters. In the most severe cases, the uncertainties in log $g$, [Fe/H], and $v_t$ produce an error in A(Li) of the order of 0.005 dex, and our largest uncertainty in $T_{eff}$, 20 K, translates to an error of 0.02 dex in A(Li).

To find the NLTE Li abundance, we performed a correction based on the median INSPECT data base[8] (Lind, Asplund & Barklem 2009). The median NLTE correction is 0.07 dex. The uncertainties of the atmospheric parameters do not seem to affect the size of the correction. Our largest uncertainties in effective temperature, log $g$, and [Fe/H] changed the NLTE Li abundances in ~0.002 dex, much lower than the typical Li error. The NLTE Li abundances, masses, and ages of our sample stars are presented in Table A1. The results for the stars with high RUWE are shown in Table A2.

## 3 RESULTS AND DISCUSSION

In general, our derived atmospheric parameters present reasonable agreement with previous results in the literature for large surveys of Li in solar-type stars (which did not have yet available the very precise Gaia DR3 parallaxes), in particular, with the results of Delgado Mena et al. (2014), with average differences in $T_{eff}$, log $g$, [Fe/H], mass, age, and A(Li) of only 8 K, 0.00 dex, 0.010 dex, 0.009 $M_\odot$, 0.9 Gyr, and 0.03 dex, respectively. More details and further comparisons are discussed in Appendix B.

---

[8]INSPECT data base: http://www.inspect-stars.com/.

### 3.1 General trends

Fig. 2 presents the Li abundance versus age for our entire sample. The solar Li abundance (A(Li)$_\odot$ = $1.07^{+0.03}_{-0.02}$ dex) is taken from Carlos et al. (2019), and was determined following similar procedures as in our work. As verified by previous studies (Monroe et al. 2013; Carlos et al. 2016; Yana Galarza et al. 2016; Carlos et al. 2019; Mishenina et al. 2020; Boesgaard et al. 2022), there is a strong correlation between A(Li) and age, with older stars exhibiting clear signs of Li depletion.

The work of Carlos et al. (2019) found a linear correlation between A(Li) and age with a dependence of $10\sigma$ for solar twin stars. Our sample, which expanded the coverage in mass for low-mass stars, demonstrates that this behaviour is more complex, showing the overall trend of Li depletion with age, but with a larger scatter than for solar twins. In particular, low-mass stars appear more Li depleted, which indicates that stellar mass is an important parameter on the Li abundance and on unravelling the nature of the non-standard mixing process.

In order to evaluate the dependence of age in A(Li) for solar twins, we selected stars with masses between 0.98 and 1.02 $M_\odot$ (same range of mass adopted by Carlos et al. 2019) from our combined sample and performed an orthogonal distance regression using the PYTHON package SCIPY.ODR. We excluded outliers (defined as stars with unusually high Li abundances; see text below) and stars with only upper limits determined for Li abundance. The resulting fit is presented in Fig. 3.

We found a linear correlation between A(Li) and age of the form

$$A(Li) = (-0.267 \pm 0.019) \times Age + (2.65 \pm 0.11). \quad (3)$$

This result reinforces what was reported by Carlos et al. (2019), in which the slope of the fit was of $-0.20 \pm 0.02$. The present work using an enlarged sample, however, shows a more powerful A(Li)-age correlation (14-$\sigma$).

A few stars in Fig. 3 present Li abundances much lower than stars with similar age, differing significantly from the fit. Nevertheless, these stars are more rich in [Fe/H], and therefore have deeper convective zones, showing enhanced Li depletion (Martos et al. 2023). Although the sample stars have metallicity close to solar, in order to exclude potential metallicity effects, we re-fitted the line considering stars with an additional restriction in metallicity ($-0.10$ dex $\leq$ [Fe/H] $\leq 0.10$ dex), but the variation in the slope of the relationship was negligible.

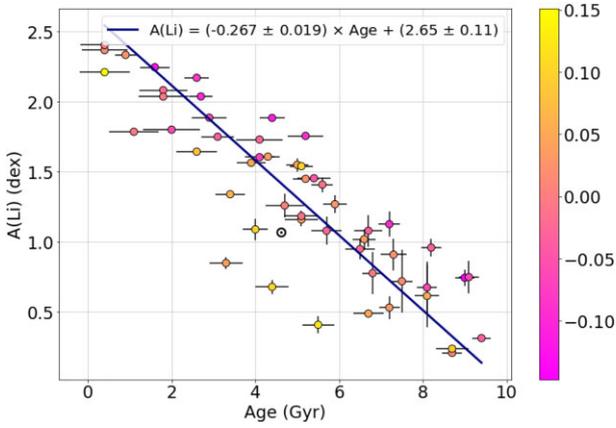

**Figure 3.** Li abundance as a function of age for solar twin stars ($0.98 \leq$ M/M$_\odot \leq 1.02$) from our combined sample, colour-coded by [Fe/H].

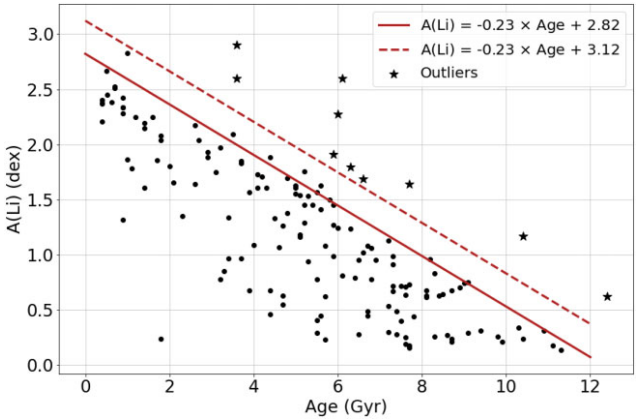

**Figure 4.** Definition of the upper envelope of A(Li) as a function of age (solid line) and outliers cut (dashed line).

Due to the dependence of metallicity on the size of the convection envelope, we would expect to see a strong connection between A(Li) and [Fe/H]. As shown in Fig. 2, metal-rich stars have in general a lower Li content, while stars with lower metallicity have somewhat higher Li abundances. However, in our sample with a narrow interval of metallicities, this effect is weak and no definite conclusions can be drawn. The influence of metallicity is more clearly outlined in the work of Martos et al. (2023).

Some stars in Fig. 2 seem to have higher Li abundances than expected for their age. In order to verify how these stars would behave in further analyses, we defined these outliers by fitting a line to a few representative stars of the 'upper envelope' of the plot, which were selected through visual inspection, taking a precaution to select at least one star from each 2 Gyr bin. The 'upper envelope' is the solid line in Fig. 4. We considered outliers the stars that present an excess of at least 0.3 dex (a factor of 2) with respect to the fitted line. This is represented by the dashed line in Fig. 4.

Fig. 5 shows the Li abundance versus mass, colour-coded by age, for the combined sample. From this figure, it is visible that as a general rule low-mass stars are more Li-poor in comparison with solar-mass stars, which is in line with our expectations since low-mass stars present deeper (therefore more massive) convective envelopes, and thus the Li can be taken towards hotter regions, where it is depleted. The lower Li abundances for stars with lower masses (or effective temperatures) is a well-known phenomenon, that was observed in many different works (mainly in open clusters), such as Zappala (1972), Soderblom et al. (1993), Sestito & Randich (2005), Pace et al. (2012), Castro et al. (2016), Carlos et al. (2016; 2019), Dumont et al. (2021), and Borisov et al. (2022). It is worth pointing out that it is possible that some of the Li destruction has occurred in the pre-main-sequence phase, when the convective zones of the stars are deeper. Thus, the non-standard mixing effects acting during the main sequence may not account for all the depleted Li.

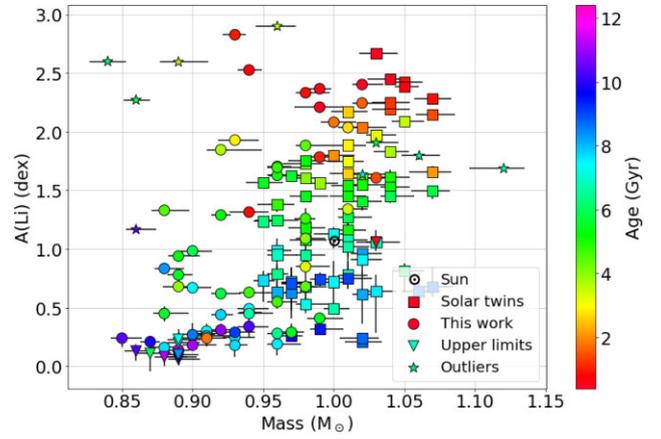

**Figure 5.** A(Li) versus mass for the stars in our sample and the solar twins from Carlos et al. (2019), colour-coded by age. The outliers of Fig. 4 are shown by stars.

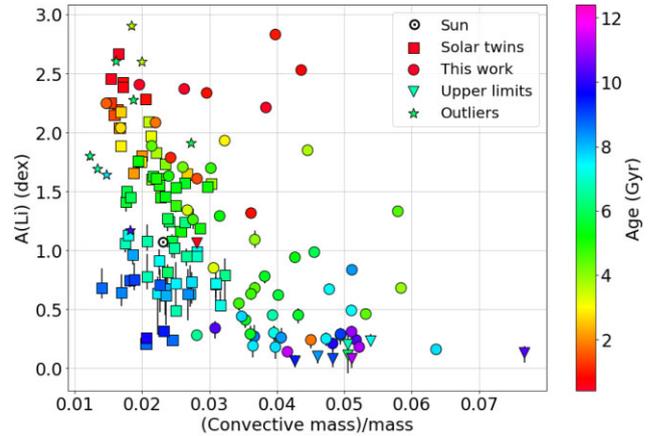

**Figure 6.** A(Li) versus (convective mass)/mass for the stars in our sample and the solar twins from Carlos et al. (2019), colour-coded by age. The outliers of Fig. 4 are shown by stars.

Similarly, Fig. 6 displays the A(Li) versus (convective mass)/mass, colour-coded by age. This figure shows that usually high-convective mass stars have lower Li abundances, as expected. Nevertheless, despite the visible effect of mass and convective mass on the Li abundance, it is clear that the correlation with age is much stronger. These results lead us to believe that although less-massive stars deplete Li more efficiently, the additional non-standard mixture process reaches deep enough in all stars in our range of masses to result in a powerful age dependence on Li destruction.

To better evaluate the effects of mass and convective mass on the Li abundance, we fitted the residuals of A(Li) as predicted by equation (3) as a function of mass and (convective mass)/mass, excluding outliers and stars with upper limits for the Li abundance. Fig. 7 presents the residuals versus mass fit, where we found a correlation

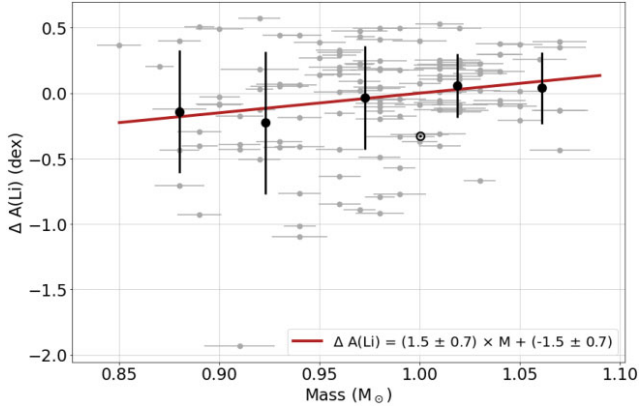

**Figure 7.** Residuals of A(Li) as a function of mass for our combined sample. Gray symbols represent our entire data set, while black symbols show the average mass and Li residual for each 0.05 M⊙ bin.

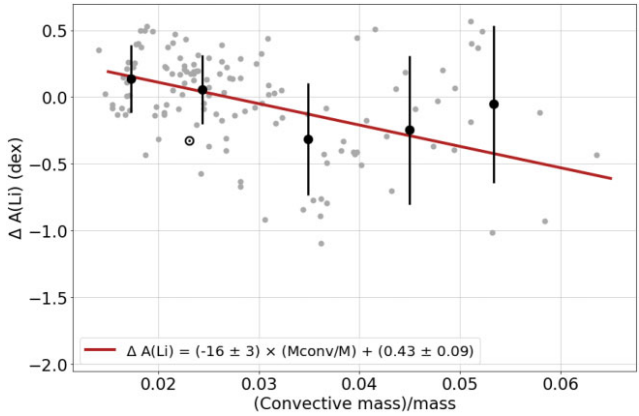

**Figure 8.** Residuals of A(Li) as a function of (convective mass)/mass for our combined sample. Gray symbols represent our entire data set, and black symbols show the average (convective mass)/mass and Li residual for each 0.01 bin.

in the form

$$\Delta A(\text{Li}) = (1.5 \pm 0.7) \times M + (-1.5 \pm 0.7). \quad (4)$$

Fig. 8 shows the residuals versus (convective mass)/mass fit, with a correlation given by

$$\Delta A(\text{Li}) = (-16 \pm 3) \times (M_{\text{conv}}/M) + (0.43 \pm 0.09). \quad (5)$$

Equations (4) and (5) reveal that the convection zone mass is a much more important factor than stellar mass for A(Li) (dependence of around 2σ for mass and about 5σ for convective mass/mass). This reflects the fact that besides mass, [Fe/H] also plays a role in the size of the convection region, thus affecting the depth that Li can reach in the stellar interior. We also fitted the Li residuals as a function of convective mass, and found a slightly less significant correlation (4σ).

There are a few outliers that have a much higher Li abundance than expected considering their age, mass, and convective mass. These stars lie in the upper-left side of Figs 5 and 6. This unusual Li abundance could be the result of a recent planet engulfment event (Sandquist et al. 2002; Meléndez et al. 2017). According to Théado & Vauclair (2012), thermohaline mixing should dilute the engulfment signature in about ∼50 Myr. Therefore, if the observed Li overabundance is due to a planet engulfment, this event must have happened on the order of a few 10 Myr ago, and it would be very unlikely to observe four such events in a sample of 151 stars, considering a typical age of 5 Gyr on the main sequence. However, Sevilla, Behmard & Fuller (2021) showed via simulations that, despite the thermohaline mixing, engulfment signatures in G-type stars (such as solar-type stars) can be detectable for ≳1 Gyr (in their work, they adopted 0.05 dex as the detection threshold). Additional tests could be performed to verify the engulfment hypothesis, such as evaluating the abundance of other refractory materials (which are richly present in rocky planets) and ⁶Li, which is a strong pollution indicator after the first few Myr following stellar formation (Sandquist et al. 2002).

In general, the stars in our sample seem to follow reasonably well the A(Li)-age correlation, with additional depletion in low-mass stars. The typical scatter of Li abundance, as estimated by the standard deviation of A(Li) in 1 Gyr bins (excluding outliers and stars with only upper limits determined), is 0.36 dex.

## 3.2 Planet connection

To verify which stars from our sample were planet hosts, we checked the NASA Exoplanet Archive data base[9] and found 15 planet hosts in our sample and six planet hosts (+the Sun) from the sample of solar twins from Carlos et al. (2019). We cannot strongly affirm that the stars with no planets reported do not host any planet; however, by construction of the sample (gathering of multi-epoch spectra to achieve a combined high SNR spectrum), our stars have time series HARPS/ESO spectra, and thus they have likely been searched for planets by previous studies.

In order to investigate the possible effect of the presence of planets on the Li abundance, we selected non-planet host stars within 0.5 Gyr, 0.1 M⊙, and 0.2 dex from the age, mass, and [Fe/H] of each planet host (similar to the bins considered by Martos et al. 2023) and obtained the average and median Li abundance from these stars. We added the sample from Martos et al. (2023), which covered a wider range of metallicity. This sample was analysed through the same high-precision methods described in this paper. Our resulting sample is composed of 192 stars, in which 35 plus the Sun have confirmed orbiting planets. The deviation of the planet host's Li abundance from the median and average A(Li) of stars without planets is presented in Fig. 9. Both the mean and the median difference (around −0.215 dex) are much larger than the typical A(Li) scatter (0.04 dex).

Considering that there should be no differences in abundances as the null hypothesis, our t-test shows that planet host stars are systematically more Li-poor in comparison with twin stars, with a significance of 99.3 per cent in the analysis and with the average of 99.6 per cent in the analysis with the median. Following Martos et al. (2023), we also tested different bins in age (1 Gyr), mass (0.05 M⊙), and metallicity (0.1 and 0.15 dex), and obtained similar results. The minimum significance found was of 82.9 per cent in the analyses with the average and 92.2 per cent in the analyses with the median. Fig. 10 displays the distribution of the significances calculated in the tests, which shows that all but one test (carried out with considering bins of) provide a significance above 90 per cent.

One possibility that could explain the systematically lower Li abundance in planet hosts is that these stars tend to be more metal-rich and thus deplete more Li due to their deeper convection zones (Martos et al. 2023). To test this hypothesis, we compared the [Fe/H]

[9]NASA Exoplanet Archive data base: https://exoplanetarchive.ipac.caltech.edu/.

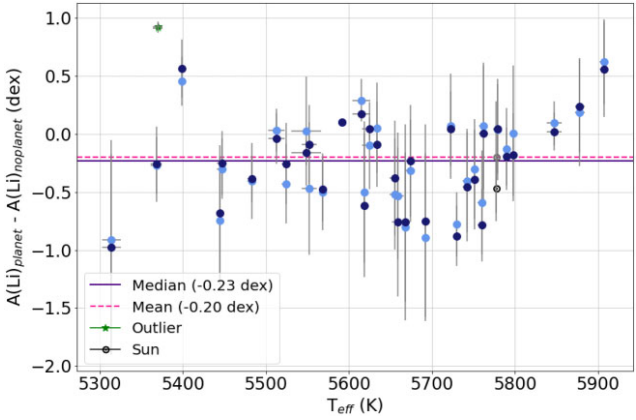

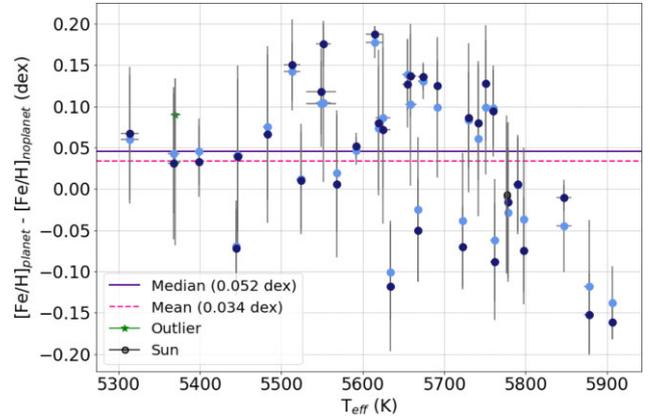

**Figure 9.** Difference between the Li abundance of planet host stars and the median (dark blue) or average (light blue) Li abundance of stars without planets of similar parameters (age ± 0.5 Gyr; mass ± 0.1 M$_\odot$; [Fe/H] ± 0.2 dex) for the stars in our sample, the solar twins from Carlos et al. (2019), and the sample from Martos et al. (2023). Error bars on the *y*-axis represent the standard deviation of the Li abundance of stars without planets for points calculated using the average A(Li) of non-planet hosts or the MAD (median absolute deviation) equivalent of the standard deviation ($\sigma_{MAD} = 1.4826 \times$ MAD) for points calculated using the median A(Li) of stars without planets. The solid line indicates the median difference using the median A(Li) of non-planet hosts and the dashed line is the average difference adopting the average A(Li) of stars without planets.

**Figure 11.** Analog to Fig. 9 for the case of [Fe/H].

### 3.3 Comparison with non-standard models

Xiong & Deng (2009) developed non-standard models for the evolution of the surface Li abundance in solar-type stars that depend on the stellar age and mass. The models start at the ZAMS (thus they do not consider PMS effects). They were tested with Li data from open clusters, and they reproduce fairly well the behaviour of Li depletion with age in young clusters such as α Per or the Pleiades (50 Myr and 70–100 Myr, respectively; Xiong & Deng 2009). In intermediate-age clusters (100 Myr–1 Gyr), the agreement is lower, but still reasonable. Fig. 12 presents a comparison of the behaviour of our sample stars and the models for different masses that consider the convective overshooting mechanism. Fig. 13 presents a comparison with the model that also takes into account gravitational settling. Notice that different masses are considered for the models on Figs 12 and 13.

In both cases, we see that the models for $M > 1$ M$_\odot$ are overabundant in Li, whereas the models for $M < 1$ M$_\odot$ burn Li too fast. The models for solar mass reproduce fairly well the behaviour seen for the solar twins, with little difference between models. However, the effect of stellar mass on the evolution of surface Li abundance is much lighter than what is predicted by Xiong & Deng (2009).

Unfortunately, the models were only evolved up to 6 Gyr, and we cannot evaluate the evolution of Li abundance during the entire main sequence (up to about 10–12 Gyr).

Another model for the dependence of A(Li) with age and mass was developed by Castro et al. (2016). Their models include the effects of convective overshooting, microscopic diffusion, mixing due to rotation, and the presence of a tachocline. The model do not consider PMS effects, but the model for the Sun was calibrated to obtain the solar A(Li) at the solar age, adopting the meteoritic abundance (from Asplund et al. 2009) as the initial abundance. Fig. 14 presents a comparison between the predictions of Castro et al. (2016) for different clusters and the stars in our sample, restricted to 1 Gyr around the age of each cluster. Since our stars present a small range of metallicity and the considered clusters have approximately solar [Fe/H] (+0.13 dex for the Hyades, 0.00 dex for NGC 752, and +0.01 dex for M67; Castro et al. 2016), we did not make a restriction in metallicity for the comparison stars.

Fig. 14 shows that although there is a reasonably good agreement between the predicted Li abundance and the measured A(Li) for supersolar masses, the models fail to explain the Li abundances for $M \lesssim 1$ M$_\odot$. Again, the observed A(Li) is larger than predicted for sub-solar masses.

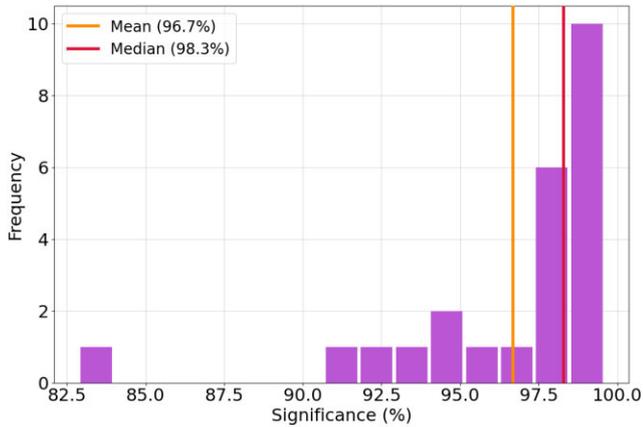

**Figure 10.** Distribution of the significances found in the tests comparing the Li abundance of stars with and without confirmed orbiting planets. Solid lines show the average (orange) and median (dark pink) significances.

of planet hosts with the average and median [Fe/H] of stars without planets of similar parameters, like in the case of Li. Our results for the analysis with the average and the median values are presented in Fig. 11. The figure reveals that planet hosts are around 0.04 dex more abundant in Fe (as measured by [Fe/H]) than non-planet hosts. Interestingly, according to equation (4) of Martos et al. (2023), a higher iron abundance by +0.04 dex would imply in a lower Li abundance by about 0.1 dex. Hence, the slightly larger metallicity of planet hosts could partly explain their lower Li abundances, but is not capable of explaining the full effect.

Therefore, our results point towards a possible correlation between the presence of planets and a low Li abundance, and calls for further studies to better investigate this connection.

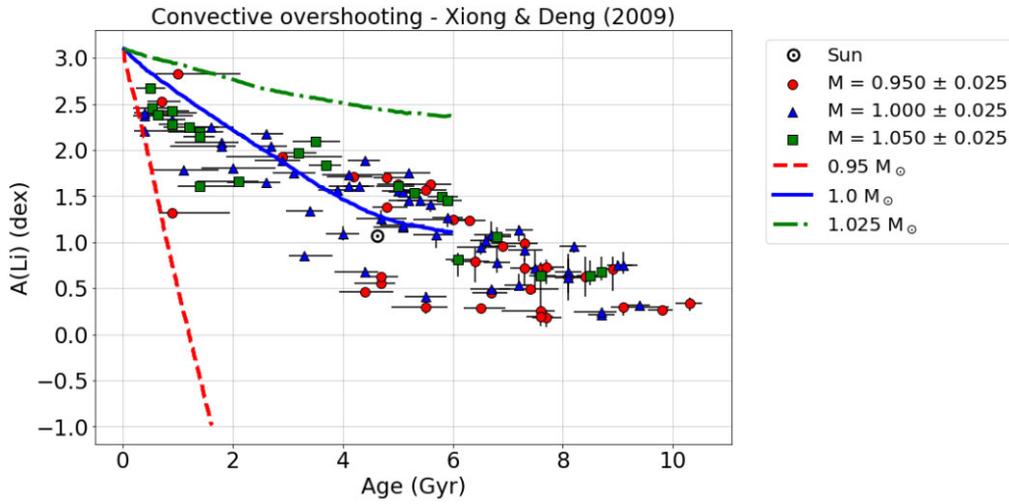

**Figure 12.** Comparison between our sample stars (filled symbols, colour-coded by mass) and the model of Xiong & Deng (2009) that considers solely convective overshooting effects (lines).

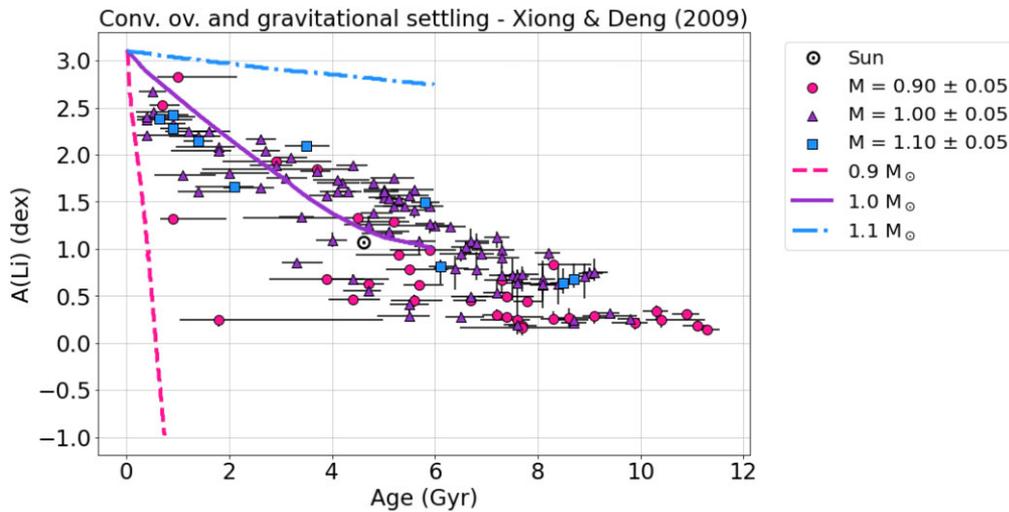

**Figure 13.** Comparison between our sample stars (filled symbols, colour-coded by mass) and the model of Xiong & Deng (2009) that considers convective overshooting and gravitational settling effects (lines).

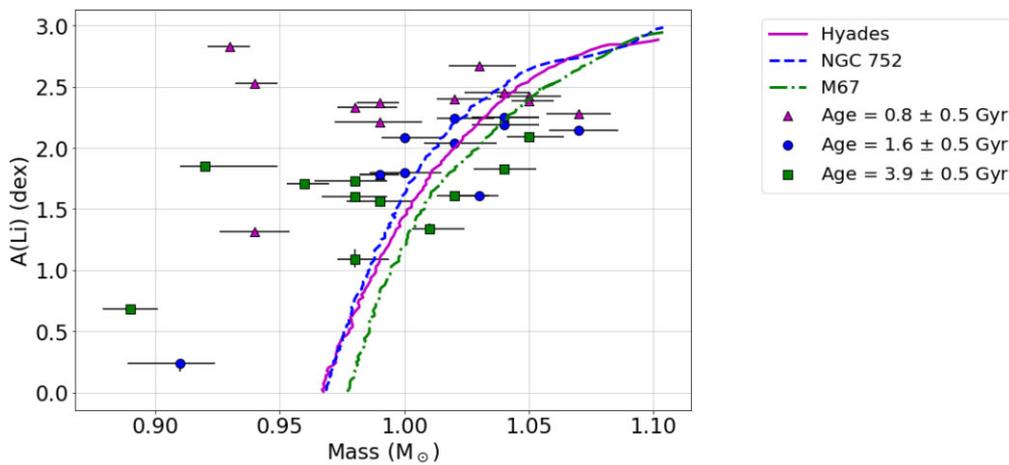

**Figure 14.** Comparison between our sample stars (filled symbols, colour-coded by age) and the models of Castro et al. (2016) for open clusters of different ages (lines).

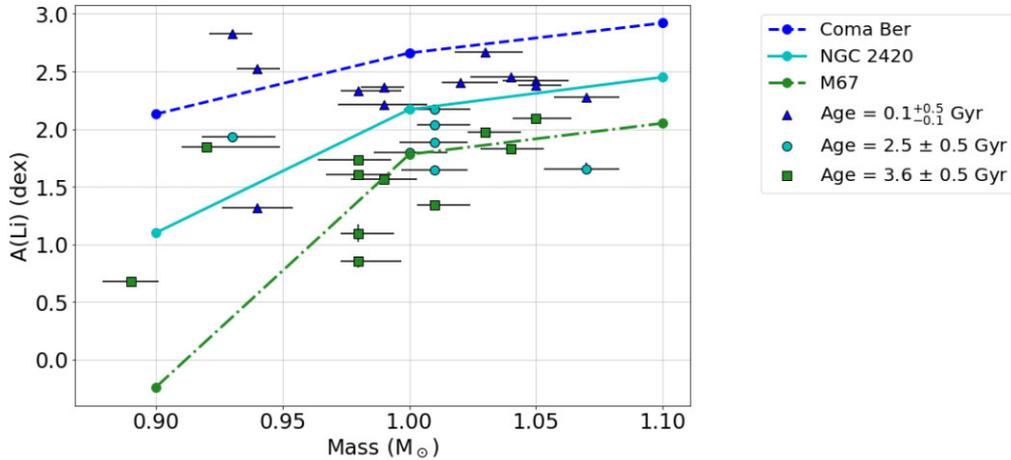

**Figure 15.** Comparison between our sample stars (filled symbols, colour-coded by age) and the models of Dumont et al. (2021) for open clusters of different ages (lines).

Finally, Dumont et al. (2021) developed age- and mass-dependent models for A(Li) in different clusters taking into account atomic diffusion, rotation-induced mixing, penetrative convection, parametric viscosity, parametric turbulence, and magnetic braking. The authors considered PMS Li effects and found that the amount of PMS Li burning depends on the stellar mass, with less-massive stars depleting more Li.

The model predictions were compared with data from different clusters, and there is excellent agreement for clusters up until ∼90 Myr. For some of the older clusters, there are important deviations from the data. Fig. 15 is the counterpart of Fig. 14 for the models of Dumont et al. (2021). Again, the considered clusters present about solar [Fe/H] (0.00 dex for Coma Ber, −0.05 dex for NGC 2420, and −0.01 dex for M67; Dumont et al. 2021), and no restrictions were made in our sample stars besides the selection in age.

The models of Dumont et al. (2021) (Fig. 15), the ones that consider the highest number of non-standard effects for A(Li), are the closest to our observations, which indicates that there are many different transport mechanisms at play in the interior of stars that must be taken into account. This highlights the complexity of the question of the mixing of material in solar analogs and the impact on the depletion of light elements in these stars.

## 4 CONCLUSIONS

Our main goal was to find correlations between lithium abundance, mass, and age in a sample of 151 solar analog stars. This new data can be used to constrain and test non-standard stellar interior models, thus helping us further understand the transport mechanisms inside stars.

We measured high-precision lithium abundances for our sample via spectral synthesis, adopting high-resolution and high-SNR HARPS spectra. The atmospheric parameters were determined employing the line-by-line differential method, and the masses and ages were found through isochronal fitting.

We confirm the strong link between Li abundance and age found in previous works. Although we find a dependence of mass ($2\sigma$) and (convective mass)/mass ($5\sigma$) on the Li abundance (with less massive stars and stars with deeper convective zones being more Li-depleted), these effects are much less important than the Li abundance–age correlation ($14\sigma$ for solar twins).

Four of our stars (HD 017925, HD 061005, HD 010008, and BD-120243) seem considerable overabundant in lithium in different plots, which is a possible evidence that they underwent a planet engulfment event. However, more investigations are required before making a definitive statement, especially considering the time-scales involved.

Thirty-six stars from our combined sample (including the stars from Carlos et al. 2019 and Martos et al. 2023) are planet hosts, from a total sample of 192 stars. We find that the planet host stars are more Li depleted by about –0.23 dex, on average, with a significance above 99 per cent for our results. Nevertheless, this result may be partly due to the somewhat higher iron abundances of planet–host stars, resulting in deeper convection zones and in stronger Li depletion (Martos et al. 2023).

We tested the mass-dependent models of Xiong & Deng (2009), Castro et al. (2016), and Dumont et al. (2021) in our sample stars, and found that the models of Xiong & Deng (2009) and Castro et al. (2016) deplete Li too fast for low-mass stars. Additionally, the models of Xiong & Deng (2009) destroy Li too slowly for stars with supersolar mass. However, the models for 1 solar mass are reasonably compatible with the behaviour of solar twins. The models of Dumont et al. (2021), that consider a wide range of non-standard effects, reproduce fairly well the observed Li abundance when tested up to about 4 Gyr.

Our results reinforce the need for further studies towards understanding the complex transport mechanisms of material in the stellar interior.

## ACKNOWLEDGEMENTS

This study was financed in part by Coordenação de Aperfeiçoamento de Pessoal de Nível Superior – Brasil (CAPES), under processes no. 88887.684392/2022-00 and 88887.823858/2023-00, and by Fundação de Amparo à Pesquisa do Estado de São Paulo – Brasil (FAPESP), under processes no. 2018/04055-8, 2019/19208-7, and 2020/15789-2. JM thanks Conselho Nacional de Desenvolvimento Científico e Tecnológico (CNPq, research fellowship 306730/2019-7).

Based on observations collected at the European Southern Observatory under ESO programmes 60.A-9036(A), 60.A-9700(G), 072.C-0488(E), 072.C-0513(D), 074.C-0012(B), 074.C-0037(A), 074.C-0364(A), 074.D-0131(E), 075.C-0202(A), 075.C-0332(A),

075.D-0194(A), 076.C-0878(B), 076.D-0130(E), 077.C-0101(A), 077.C-0364(E), 077.C-0530(A), 078.D-0071(E), 080.D-0086(E), 081.D-0065(A), 082.C-0390(A), 082.C-0427(A), 085.C-0019(A), 087.C-0831(A), 089.C-0732(A), 090.C-0421(A), 091.C-0034(A), 091.C-0936(A), 092.C-0721(A), 095.C-0551(A), 095.D-0717(A), 096.C-0460(A), 096.C-0499(A), 097.C-0090(A), 097.D-0717(A), 098.C-0739(A), 099.C-0458(A), 0100.C-0097(A), 0100.C-0414(B), 0100.C-0474(A), 0100.C-0487(A), 0100.D-0444(A), 0101.C-0232(A), 0101.C-0232(B), 0101.C-0379(A), 0102.C-0338(A), 0102.C-0558(A), 0102.C-0584(A), 0103.C-0206(A), 0103.C-0432(A), 0103.C-0548(A), 0104.C-0090(A), 0104.C-0090(B), 105.20AK.002, 106.215E.002, 109.2374.001, 1102.D-0954(A), 183.C-0972(A), 184.C-0815(B), 185.D-0056(F), 185.D-0056(K), 192.C-0224(A), 192.C-0224(C), 192.C-0224(G), 192.C-0852(A), 196.C-1006(A), and 198.C-0836(A).

AR would like to acknowledge the support from the members of the SAMPA (Stellar Atmospheres, Planets and Abundances) group at IAG-USP, and the important reviews from Francisco G. Barbosa and Thiago F. dos Santos.

## DATA AVAILABILITY

The spectra used in this work are publicly available at ESO's online data base (ESO Science Archive Facility, http://archive.eso.org/).

## APPENDIX A: SAMPLE STELLAR PARAMETERS AND LITHIUM ABUNDANCES

**Table A1.** Atmospheric parameters, Li abundances, ages, and masses for the solar analogues.

| HIP | Other ID | $T_{eff}$ (K) | log $g$ (dex) | [Fe/H] (dex) | $v_t$ (km s$^{-1}$) | LTE A(Li) (dex) | NLTE A(Li) (dex) | Age (Gyr) | Mass ($M_\odot$) | $M_{conv.}$ ($M_\odot$) |
|---|---|---|---|---|---|---|---|---|---|---|
| 000669 | HD 000361 | 5902 ± 7 | 4.520 ± 0.013 | −0.122 ± 0.005 | 1.10 ± 0.01 | 2.240 | 2.245 ± 0.013 | $1.60^{+0.42}_{-0.30}$ | $1.02^{+0.02}_{-0.01}$ | 0.015 |
| 001292 | HD 001237 | 5549 ± 18 | 4.550 ± 0.031 | 0.151 ± 0.015 | 1.10 ± 0.04 | 2.140 | $2.210^{+0.023}_{-0.022}$ | $0.40^{+0.39}_{-0.26}$ | 0.99 ± 0.02 | 0.038 |
| 003311 | HD 003964 | 5734 ± 3 | 4.47 ± 0.01 | 0.064 ± 0.003 | 0.96 ± 0.01 | 1.290 | $1.340^{+0.041}_{-0.040}$ | $3.40^{+0.41}_{-0.31}$ | 1.01 ± 0.01 | 0.027 |
| 005938 | HD 007661 | 5450 ± 14 | 4.520 ± 0.024 | 0.032 ± 0.010 | 1.08 ± 0.03 | 1.770 | 1.847 ± 0.019 | $3.70^{+4.41}_{-0.71}$ | $0.92^{+0.03}_{-0.02}$ | 0.041 |
| 006276 | BD −12 0243 | 5384 ± 18 | 4.54 ± 0.03 | −0.014 ± 0.012 | 1.14 ± 0.03 | 2.590 | $2.596^{+0.023}_{-0.021}$ | $3.60^{+1.38}_{-1.33}$ | $0.89^{+0.02}_{-0.01}$ | 0.044 |
| 006455 | HD 008406 | 5731 ± 4 | 4.480 ± 0.013 | −0.105 ± 0.003 | 0.96 ± 0.01 | 1.670 | 1.705 ± 0.011 | $4.20^{+0.49}_{-0.33}$ | 0.96 ± 0.01 | 0.025 |
| 006744 | HD 008859 | 5518 ± 4 | 4.390 ± 0.013 | −0.069 ± 0.004 | 0.82 ± 0.01 | 0.23 | $0.30^{+0.07}_{-0.07}$ | $7.20^{+0.29}_{-0.28}$ | 0.91 ± 0.01 | 0.036 |
| 006762 | HD 008828 | 5398 ± 6 | 4.388 ± 0.019 | −0.142 ± 0.005 | 0.72 ± 0.02 | 0.16 | $0.24^{+0.04}_{-0.04}$ | $10.40^{+0.30}_{-0.38}$ | 0.85 ± 0.01 | 0.044 |
| 007576 | HD 010008 | 5316 ± 9 | 4.490 ± 0.024 | −0.055 ± 0.008 | 0.93 ± 0.03 | 2.220 | $2.272^{+0.010}_{-0.010}$ | $6.00^{+0.64}_{-0.62}$ | 0.86 ± 0.01 | 0.048 |
| 009400 | HD 013060 | 5231 ± 11 | 4.36 ± 0.03 | 0.09 ± 0.01 | 0.68 ± 0.04 | 0.04 | 0.16 ± 0.05 | $7.70^{+1.08}_{-1.42}$ | 0.88 ± 0.01 | 0.056 |
| 010818 | HD 014374 | 5413 ± 6 | 4.450 ± 0.022 | −0.018 ± 0.006 | 0.81 ± 0.02 | 0.900 | $0.984^{+0.012}_{-0.011}$ | $5.90^{+0.78}_{-0.92}$ | 0.90 ± 0.01 | 0.041 |
| 012119 | HD 016297 | 5396 ± 6 | 4.430 ± 0.019 | 0.008 ± 0.006 | 0.81 ± 0.02 | <0.11 | $<0.20^{+0.02}_{-0.05}$ | $6.80^{+2.59}_{-1.46}$ | $0.89^{+0.02}_{-0.01}$ | 0.045 |
| 013402 | HD 017925 | 5149 ± 18 | 4.435 ± 0.057 | 0.088 ± 0.015 | 0.97 ± 0.06 | 2.60 | 2.61 ± 0.02 | $6.10^{+2.58}_{-2.36}$ | 0.84 ± 0.01 | 0.066 |
| 014530 | HD 020003 | 5483 ± 5 | 4.400 ± 0.019 | 0.075 ± 0.006 | 0.76 ± 0.02 | 0.17 | $0.25^{+0.05}_{-0.06}$ | $7.60^{+0.26}_{-0.35}$ | 0.93 ± 0.01 | 0.044 |
| 014684 | HD 019668 | 5501 ± 18 | 4.570 ± 0.034 | 0.010 ± 0.013 | 1.15 ± 0.04 | 2.870 | 2.828 ± 0.024 | $1.00^{+1.13}_{-0.60}$ | 0.93 ± 0.01 | 0.037 |
| 015799 | HD 021175 | 5223 ± 12 | 4.348 ± 0.034 | 0.127 ± 0.010 | 0.69 ± 0.04 | <0.00 | $<0.13^{+0.06}_{-0.08}$ | $10.30^{+1.02}_{-1.17}$ | 0.86 ± 0.01 | 0.066 |
| 016085 | HD 021693 | 5447 ± 6 | 4.410 ± 0.019 | 0.044 ± 0.006 | 0.73 ± 0.02 | 0.19 | $0.28^{+0.08}_{-0.08}$ | $7.40^{+0.40}_{-0.37}$ | 0.91 ± 0.01 | 0.045 |
| 019855 | HD 026913 | 5692 ± 14 | 4.560 ± 0.027 | 0.015 ± 0.010 | 1.20 ± 0.03 | 2.340 | $2.367^{+0.010}_{-0.010}$ | $0.40^{+0.88}_{-0.38}$ | 0.99 ± 0.01 | 0.026 |
| 022504 | HD 034449 | 5854 ± 3 | 4.47 ± 0.01 | −0.098 ± 0.003 | 1.07 ± 0.01 | 2.020 | 2.037 ± 0.011 | $2.70^{+0.35}_{-0.45}$ | 1.01 ± 0.01 | 0.017 |
| 025191 | HD 290327 | 5524 ± 5 | 4.420 ± 0.016 | −0.103 ± 0.006 | 0.78 ± 0.02 | <0.00 | <0.06 ± 0.05 | $9.40^{+0.84}_{-0.54}$ | $0.89^{+0.02}_{-0.01}$ | 0.038 |
| 028267 | HD 040397 | 5522 ± 5 | 4.380 ± 0.014 | −0.116 ± 0.004 | 0.84 ± 0.01 | 0.08 | 0.14 ± 0.05 | $11.30^{+0.25}_{-0.25}$ | 0.89 ± 0.01 | 0.037 |
| 028818 | HD 041087 | 5588 ± 13 | 4.53 ± 0.02 | −0.105 ± 0.009 | 1.07 ± 0.02 | 1.880 | $1.929^{+0.025}_{-0.103}$ | $2.90^{+1.09}_{-1.01}$ | $0.93^{+0.02}_{-0.03}$ | 0.030 |
| 028954 | HD 041593 | 5306 ± 14 | 4.490 ± 0.031 | 0.034 ± 0.011 | 0.93 ± 0.04 | 1.23 | 1.33 ± 0.02 | $4.50^{+2.23}_{-2.33}$ | $0.88^{+0.03}_{-0.02}$ | 0.051 |
| 029271 | HD 043834 | 5593 ± 5 | 4.375 ± 0.015 | 0.123 ± 0.005 | 0.86 ± 0.01 | 0.34 | $0.41^{+0.05}_{-0.05}$ | $5.50^{+0.38}_{-0.31}$ | $0.99^{+0.02}_{-0.01}$ | 0.035 |
| 029525 | HD 042807 | 5747 ± 11 | 4.54 ± 0.019 | −0.018 ± 0.010 | 1.10 ± 0.02 | 2.050 | $2.082^{+0.016}_{-0.016}$ | $1.80^{+0.31}_{-0.31}$ | $1.00^{+0.01}_{-0.01}$ | 0.022 |
| 029568 | HD 043162 | 5658 ± 13 | 4.550 ± 0.025 | 0.025 ± 0.009 | 1.15 ± 0.02 | 2.300 | $2.333^{+0.015}_{-0.015}$ | $0.90^{+0.27}_{-0.55}$ | $0.98^{+0.01}_{-0.01}$ | 0.029 |
| 033229 | HD 051608 | 5370 ± 6 | 4.38 ± 0.02 | −0.023 ± 0.006 | 0.70 ± 0.02 | 1.080 | 1.167 ± 0.012 | $10.40^{+0.35}_{-0.22}$ | 0.86 ± 0.01 | 0.050 |
| 036210 | HD 059468 | 5617 ± 4 | 4.385 ± 0.012 | 0.045 ± 0.004 | 0.91 ± 0.01 | 0.13 | $0.19^{+0.07}_{-0.07}$ | $7.60^{+0.22}_{-0.22}$ | $0.96^{+0.02}_{-0.01}$ | 0.035 |
| 036948 | HD 061005 | 5581 ± 20 | 4.51 ± 0.04 | 0.069 ± 0.015 | 1.43 ± 0.04 | 2.94 | 2.90 ± 0.03 | $3.60^{+2.16}_{-2.16}$ | $0.96^{+0.01}_{-0.01}$ | 0.037 |
| 038041 | HD 063765 | 5445 ± 6 | 4.475 ± 0.016 | −0.137 ± 0.004 | 0.81 ± 0.01 | 0.38 | $0.45^{+0.06}_{-0.05}$ | $5.60^{+1.94}_{-1.85}$ | $0.88^{+0.02}_{-0.01}$ | 0.038 |
| 038558 | HD 065216 | 5634 ± 4 | 4.470 ± 0.013 | −0.155 ± 0.004 | 0.91 ± 0.01 | 1.24 | 1.29 ± 0.02 | $5.20^{+0.80}_{-0.73}$ | 0.92 ± 0.01 | 0.029 |
| 040693 | HD 069830 | 5399 ± 5 | 4.410 ± 0.017 | −0.029 ± 0.006 | 0.71 ± 0.02 | 0.750 | 0.834 ± 0.021 | $8.30^{+0.38}_{-0.35}$ | 0.88 ± 0.01 | 0.045 |
| 041529 | HD 071835 | 5464 ± 4 | 4.435 ± 0.014 | 0.005 ± 0.004 | 0.78 ± 0.01 | 0.59 | $0.67^{+0.04}_{-0.03}$ | $7.30^{+2.36}_{-2.36}$ | 0.90 ± 0.01 | 0.043 |
| 044890 | HD 078538 | 5804 ± 9 | 4.530 ± 0.015 | −0.006 ± 0.006 | 1.07 ± 0.01 | 2.390 | $2.403^{+0.010}_{-0.020}$ | $0.40^{+0.84}_{-0.31}$ | $1.02^{+0.02}_{-0.01}$ | 0.020 |
| 050534 | HD 089454 | 5733 ± 4 | 4.500 ± 0.013 | 0.139 ± 0.004 | 0.98 ± 0.01 | 1.550 | $1.607^{+0.011}_{-0.011}$ | $1.40^{+0.32}_{-0.20}$ | 1.03 ± 0.01 | 0.029 |
| 052369 | HD 092719 | 5812 ± 3 | 4.470 ± 0.011 | −0.114 ± 0.003 | 1.03 ± 0.01 | 1.860 | 1.884 ± 0.011 | $4.40^{+0.35}_{-0.30}$ | $0.98^{+0.02}_{-0.01}$ | 0.021 |
| 053087 | HD 094151 | 5618 ± 3 | 4.430 ± 0.012 | 0.072 ± 0.003 | 0.89 ± 0.01 | 0.22 | 0.29 ± 0.07 | $5.50^{+0.60}_{-0.62}$ | 0.97 ± 0.01 | 0.035 |

**Table A1** *– continued*

| HIP | Other ID | $T_{eff}$ (K) | log $g$ (dex) | [Fe/H] (dex) | $v_t$ (km s$^{-1}$) | LTE A(Li) (dex) | NLTE A(Li) (dex) | Age (Gyr) | Mass (M$_\odot$) | M$_{conv.}$ (M$_\odot$) |
|---|---|---|---|---|---|---|---|---|---|---|
| 053837 | HD 095521 | 5774 ± 4 | 4.480 ± 0.013 | −0.155 ± 0.004 | 1.02 ± 0.01 | 1.600 | 1.629 ± 0.011 | $5.00^{+0.29}_{-0.30}$ | $0.96^{+0.02}_{-0.01}$ | 0.023 |
| 054155 | HD 096064 | 5307 ± 17 | 4.56 ± 0.03 | 0.067 ± 0.011 | 1.16 ± 0.03 | 2.490 | 2.527 ± 0.022 | $0.70^{+0.38}_{-0.28}$ | 0.94 ± 0.01 | 0.041 |
| 054704 | HD 097343 | 5400 ± 6 | 4.350 ± 0.017 | −0.032 ± 0.006 | 0.69 ± 0.02 | <0.00 | $<0.08^{+0.04}_{-0.08}$ | $11.20^{+0.32}_{-0.28}$ | $0.88^{+0.02}_{-0.01}$ | 0.045 |
| 057370 | HD 102195 | 5314 ± 11 | 4.48 ± 0.03 | 0.073 ± 0.009 | 0.92 ± 0.03 | 0.57 | 0.68 ± 0.03 | $3.90^{+1.12}_{-1.36}$ | 0.89 ± 0.01 | 0.052 |
| 058558 | HD 104263 | 5484 ± 5 | 4.32 ± 0.02 | 0.051 ± 0.005 | 0.82 ± 0.01 | 0.10 | 0.18 ± 0.05 | $11.10^{+0.54}_{-0.27}$ | $0.90^{+0.02}_{-0.01}$ | 0.047 |
| 064109 | HD 112540 | 5526 ± 6 | 4.460 ± 0.018 | −0.164 ± 0.007 | 0.84 ± 0.02 | 0.72 | $0.78^{+0.04}_{-0.05}$ | $5.50^{+0.76}_{-1.25}$ | 0.89 ± 0.01 | 0.034 |
| 064924 | HD 115617 | 5568 ± 4 | 4.390 ± 0.012 | 0.006 ± 0.004 | 0.84 ± 0.01 | 0.11 | $0.18^{+0.08}_{-0.10}$ | $7.70^{+0.28}_{-0.44}$ | 0.93 ± 0.01 | 0.037 |
| 069414 | HD 124292 | 5458 ± 4 | 4.390 ± 0.014 | −0.113 ± 0.004 | 0.79 ± 0.01 | 0.14 | $0.21^{+0.08}_{-0.07}$ | $9.90^{+0.44}_{-0.24}$ | 0.87 ± 0.01 | 0.042 |
| 070695 | HD 126525 | 5668 ± 3 | 4.42 ± 0.01 | −0.075 ± 0.003 | 0.94 ± 0.01 | 0.23 | 0.28 ± 0.05 | $6.50^{+0.40}_{-0.46}$ | 0.96 ± 0.01 | 0.027 |
| 072339 | HD 130322 | 5368 ± 7 | 4.415 ± 0.022 | 0.036 ± 0.007 | 0.74 ± 0.02 | <0.13 | $<-0.23^{+0.06}_{-0.04}$ | $7.30^{+0.70}_{-4.30}$ | $0.89^{+0.02}_{-0.01}$ | 0.048 |
| 074271 | HD 134330 | 5623 ± 5 | 4.458 ± 0.016 | 0.102 ± 0.005 | 0.98 ± 0.01 | 1.02 | $1.09^{+0.08}_{-0.07}$ | $4.00^{+0.36}_{-0.08}$ | 0.98 ± 0.01 | 0.036 |
| 075363 | HD 136894 | 5451 ± 5 | 4.39 ± 0.02 | −0.062 ± 0.006 | 0.79 ± 0.02 | <0.02 | $<-0.10^{+0.08}_{-0.09}$ | $8.30^{+0.43}_{-0.33}$ | 0.89 ± 0.01 | 0.041 |
| 076200 | HD 138549 | 5586 ± 3 | 4.445 ± 0.012 | 0.032 ± 0.005 | 0.89 ± 0.01 | 0.48 | $0.55^{+0.06}_{-0.06}$ | $4.70^{+0.49}_{-0.43}$ | 0.96 ± 0.01 | 0.033 |
| 077358 | HD 140901 | 5619 ± 4 | 4.430 ± 0.015 | 0.107 ± 0.004 | 0.95 ± 0.01 | 0.61 | 0.68 ± 0.05 | $4.40^{+0.45}_{-0.43}$ | 0.98 ± 0.01 | 0.036 |
| 080218 | HD 147512 | 5535 ± 4 | 4.410 ± 0.013 | −0.061 ± 0.004 | 0.81 ± 0.01 | 0.19 | $0.26^{+0.08}_{-0.08}$ | $8.30^{+0.38}_{-0.35}$ | $0.91^{+0.02}_{-0.02}$ | 0.037 |
| 081300 | HD 149661 | 5248 ± 10 | 4.45 ± 0.03 | 0.012 ± 0.011 | 0.89 ± 0.04 | 0.13 | $0.24^{+0.05}_{-0.03}$ | $1.80^{+3.19}_{-1.10}$ | $0.91^{+0.01}_{-0.02}$ | 0.041 |
| 082265 | HD 151504 | 5443 ± 6 | 4.335 ± 0.019 | 0.098 ± 0.006 | 0.72 ± 0.02 | 0.20 | $0.29^{+0.07}_{-0.07}$ | $9.10^{+0.20}_{-0.63}$ | 0.93 ± 0.01 | 0.046 |
| 082588 | HD 152391 | 5483 ± 10 | 4.525 ± 0.021 | −0.003 ± 0.008 | 0.97 ± 0.02 | 1.240 | $1.315^{+0.011}_{-0.023}$ | $0.90^{+0.05}_{-0.108}$ | 0.94 ± 0.01 | 0.034 |
| 085017 | HD 157172 | 5435 ± 7 | 4.34 ± 0.02 | 0.140 ± 0.008 | 0.73 ± 0.02 | 0.39 | 0.49 ± 0.03 | $7.40^{+0.35}_{-0.53}$ | 0.94 ± 0.01 | 0.048 |
| 087369 | HD 162236 | 5365 ± 6 | 4.425 ± 0.021 | −0.076 ± 0.005 | 0.79 ± 0.02 | <0.02 | $<0.11^{+0.08}_{-0.04}$ | $6.00^{+1.01}_{-0.93}$ | $0.87^{+0.02}_{-0.01}$ | 0.044 |
| 090004 | HD 168746 | 5592 ± 5 | 4.360 ± 0.014 | −0.090 ± 0.005 | 0.90 ± 0.01 | 0.28 | $0.34^{+0.06}_{-0.09}$ | $10.30^{+0.42}_{-0.35}$ | 0.94 ± 0.01 | 0.029 |
| 091287 | HD 171665 | 5667 ± 3 | 4.440 ± 0.01 | −0.037 ± 0.003 | 0.94 ± 0.01 | 1.650 | 1.696 ± 0.011 | $4.80^{+0.25}_{-0.35}$ | $0.96^{+0.02}_{-0.01}$ | 0.029 |
| 091700 | HD 172513 | 5525 ± 4 | 4.460 ± 0.013 | −0.023 ± 0.004 | 0.86 ± 0.01 | 0.55 | 0.62 ± 0.03 | $5.70^{+0.30}_{-0.44}$ | 0.92 ± 0.01 | 0.037 |
| 097358 | HD 186803 | 5684 ± 9 | 4.540 ± 0.018 | −0.009 ± 0.007 | 0.96 ± 0.02 | 1.740 | $1.785^{+0.022}_{-0.024}$ | $1.10^{+0.56}_{-0.85}$ | 0.99 ± 0.01 | 0.024 |
| 097507 | HD 186302 | 5694 ± 6 | 4.48 ± 0.02 | 0.007 ± 0.007 | 0.92 ± 0.02 | 1.21 | $1.26^{+0.09}_{-0.09}$ | $4.70^{+0.46}_{-0.48}$ | 0.98 ± 0.01 | 0.027 |
| 098621 | HD 188748 | 5629 ± 4 | 4.420 ± 0.012 | −0.114 ± 0.004 | 0.92 ± 0.01 | 0.39 | 0.44 ± 0.05 | $7.80^{+0.30}_{-0.37}$ | 0.92 ± 0.01 | 0.032 |
| 102203 | HD 197210 | 5583 ± 3 | 4.47 ± 0.01 | −0.009 ± 0.004 | 0.85 ± 0.01 | 0.57 | $0.63^{+0.04}_{-0.04}$ | $4.70^{+0.35}_{-0.35}$ | 0.94 ± 0.01 | 0.034 |
| 102580 | HD 197823 | 5400 ± 7 | 4.400 ± 0.023 | 0.164 ± 0.007 | 0.79 ± 0.02 | 0.36 | $0.46^{+0.05}_{-0.05}$ | $4.40^{+0.40}_{-0.47}$ | 0.94 ± 0.01 | 0.050 |
| 104318 | HD 201219✱ | 5669 ± 11 | 4.54 ± 0.02 | 0.132 ± 0.008 | 1.02 ± 0.02 | <0.99 | $<1.06^{+0.04}_{-0.05}$ | 0.50 ± 0.36 | $1.030^{+0.003}_{-0.014}$ | 0.029 |
| 106931 | HD 205891 | 5574 ± 4 | 4.400 ± 0.012 | −0.138 ± 0.004 | 0.86 ± 0.01 | 0.21 | 0.27 ± 0.10 | $8.60^{+0.82}_{-0.30}$ | $0.90^{+0.02}_{-0.01}$ | 0.033 |
| 107022 | HD 205536 | 5438 ± 5 | 4.390 ± 0.017 | −0.021 ± 0.006 | 0.73 ± 0.02 | <0.00 | $<0.08^{+0.10}_{-0.07}$ | $9.00^{+0.24}_{-0.25}$ | 0.89 ± 0.01 | 0.043 |
| 108065 | HD 207970 | 5562 ± 6 | 4.32 ± 0.02 | 0.086 ± 0.005 | 0.89 ± 0.01 | 0.24 | 0.31 ± 0.05 | $10.90^{+0.25}_{-0.29}$ | 0.92 ± 0.01 | 0.047 |
| 111349 | HD 213628 | 5549 ± 5 | 4.405 ± 0.015 | 0.026 ± 0.005 | 0.81 ± 0.01 | 0.38 | 0.45 ± 0.03 | $6.70^{+0.35}_{-0.35}$ | $0.94^{+0.02}_{-0.01}$ | 0.037 |
| 116819 | HD 222422 | 5467 ± 5 | 4.450 ± 0.018 | −0.104 ± 0.006 | 0.77 ± 0.02 | 0.87 | 0.94 ± 0.03 | $5.30^{+0.35}_{-0.43}$ | $0.89^{+0.02}_{-0.01}$ | 0.038 |
| 116937 | HD 222595 | 5649 ± 3 | 4.465 ± 0.010 | 0.039 ± 0.003 | 0.92 ± 0.01 | 0.79 | $0.85^{+0.03}_{-0.05}$ | $3.30^{+0.45}_{-0.27}$ | $0.98^{+0.02}_{-0.01}$ | 0.030 |

**Table A2.** Atmospheric parameters, Li abundances, ages, and masses for the stars with high RUWE.

| HIP | Other ID | $T_{\rm eff}$ (K) | $\log g$ (dex) | [Fe/H] (dex) | $v_{\rm t}$ (km s$^{-1}$) | LTE A(Li) (dex) | NLTE A(Li) (dex) | Age (Gyr) | Mass (M$_\odot$) | $M_{\rm conv.}$ (M$_\odot$) |
|---|---|---|---|---|---|---|---|---|---|---|
| 090355 | HD 169822 | 5567 ± 9 | 4.360 ± 0.025 | −0.183 ± 0.007 | 0.90 ± 0.02 | 0.57 | 0.62 ± 0.10 | $12.40^{+0.74}_{-0.88}$ | $0.88^{+0.02}_{-0.01}$ | 0.035 |
| 088601 | HD 165341 | 5335 ± 9 | 4.45 ± 0.03 | 0.052 ± 0.009 | 0.92 ± 0.03 | 0.10 | $0.20^{+0.07}_{-0.05}$ | $10.10^{+0.86}_{-0.82}$ | 0.87 ± 0.01 | 0.055 |

# APPENDIX B: COMPARISON WITH PREVIOUS RESULTS

In this section, we compare our results for Li abundances, ages, masses, and atmospheric parameters with the ones derived in large surveys of Li in solar-type stars, namely, Ramírez et al. (2012), Delgado Mena et al. (2014), and Bensby & Lind (2018). We highlight the fact that our work benefits from high precision parallaxes from Gaia's DR3, which were not available for the previous studies.

Figs B1 and B2 show the comparison between LTE and NLTE Li abundances with the aforementioned works, respectively. Figs B3–B8 present the comparison for effective temperature, log $g$, microturbulence velocity, [Fe/H], age, and mass.

The results are summarized in Table B1, that show the average differences and their standard deviations. Our results are compatible with the results of Ramírez et al. (2012) considering ours and their typical uncertainties (44 K for $T_{\rm eff}$, 0.04 dex for log $g$, 0.05 dex for [Fe/H], and 0.04 for Li abundance; their table 1).

Our results for LTE A(Li) agree with the results of Delgado Mena et al. (2014), with the exception of HD 186302, for which their A(Li) is much lower. This star was included in the plots, but discarded when calculating the average differences between stellar parameters and their standard deviations. The microturbulence velocities presented in this work are systematically lower by around 0.073 dex. None the less, this value is about the same as their typical uncertainty (0.08 km s$^{-1}$) and, besides, our $v_{\rm t}$ are obtained

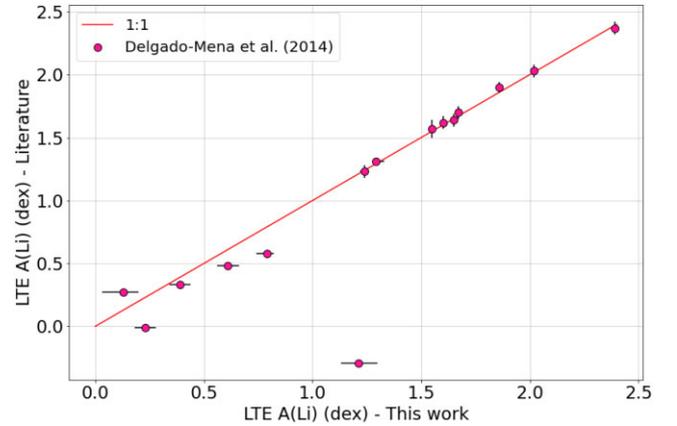

**Figure B1.** Comparison between the LTE Li abundances derived in this work and the LTE Li abundance from Delgado Mena et al. (2014).

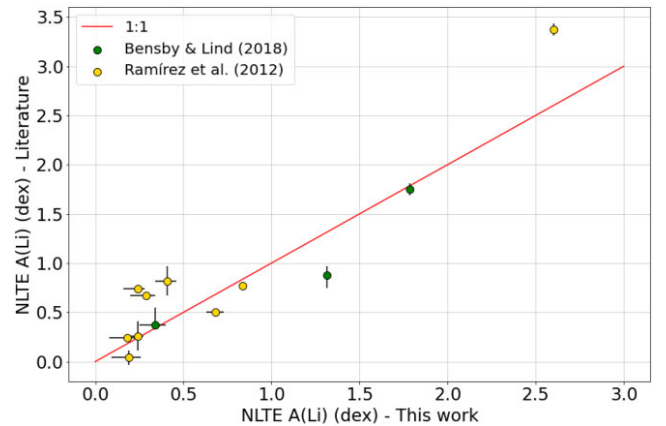

**Figure B2.** Comparison between the NLTE Li abundances determined in this work and NLTE Li abundances from the literature.

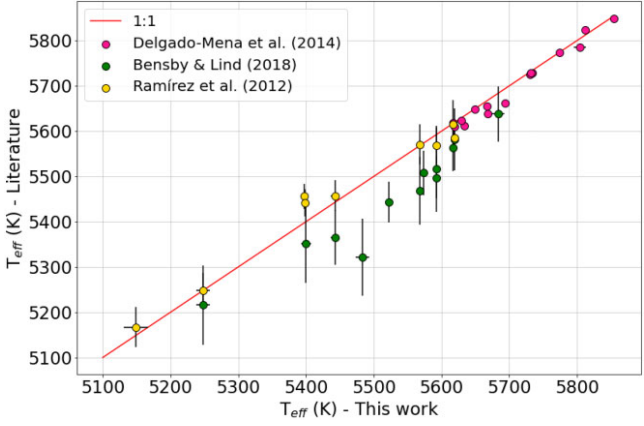

**Figure B3.** Comparison between the $T_{eff}$ derived in this work and the $T_{eff}$ from the literature.

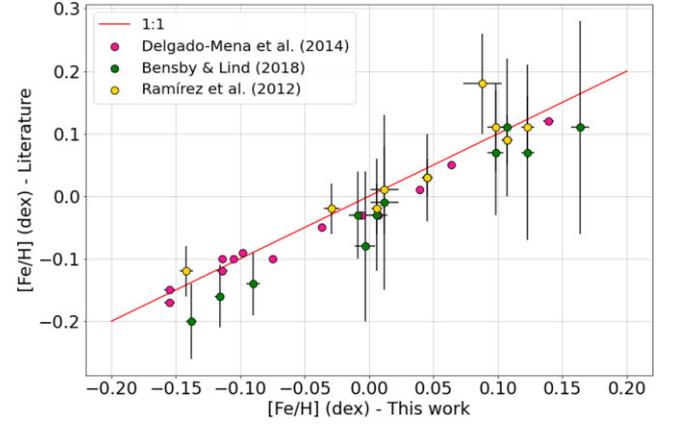

**Figure B6.** Comparison between the [Fe/H] derived in this work and the [Fe/H] from the literature.

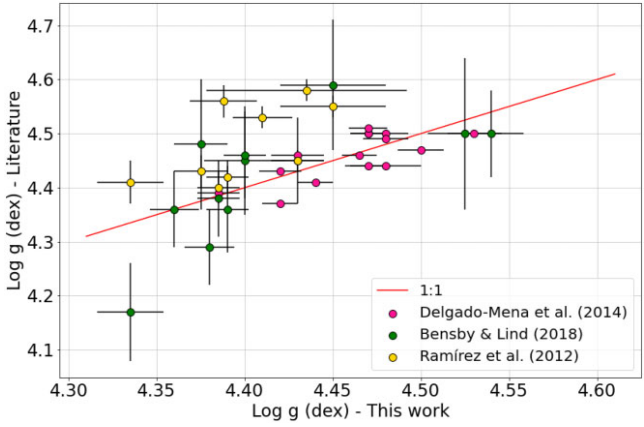

**Figure B4.** Comparison between the $\log g$ found in this work and the $\log g$ from the literature.

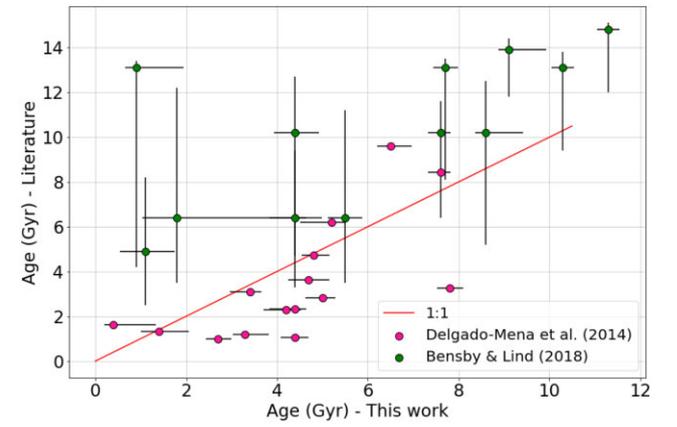

**Figure B7.** Comparison between the ages calculated in this work and the ages from the literature.

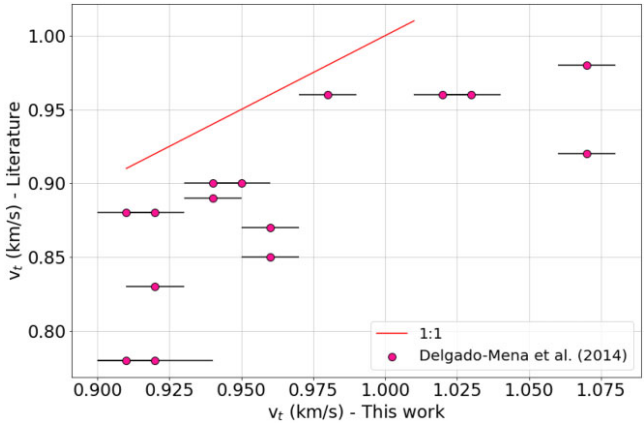

**Figure B5.** Comparison between the $v_t$ determined in this work and the $v_t$ from Delgado Mena et al. (2014).

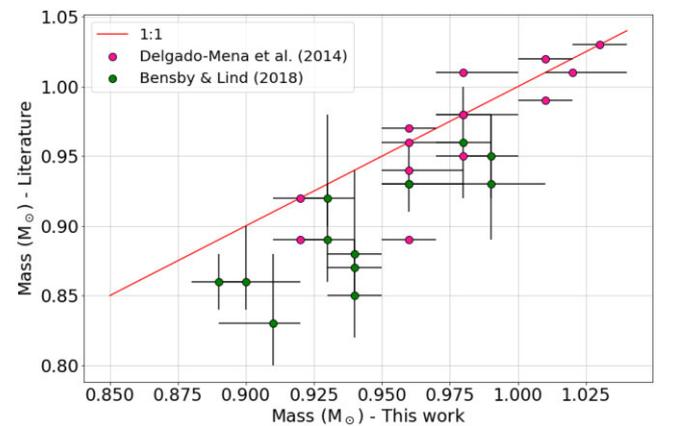

**Figure B8.** Comparison between the masses derived in this work and the masses from the literature.

relative to the adopted Sun's value. The remaining parameters are compatible with our determinations, with low average differences and low standard deviations.

Finally, the work of Bensby & Lind (2018) present parameters that agree with the ones derived in this work considering the uncertainties (their typical uncertainties are of 56 K for $T_{eff}$, 0.08 dex for $\log g$, 0.05 dex for [Fe/H], 1.8 Gyr for age, 0.04 $M_{\odot}$ for stellar mass, and 0.04 dex for A(Li); table 2 in Bensby & Lind 2018). Their ages, however, are systematically larger than ours. Nevertheless, the ages in their work were determined through the $\log g$ method using $q^2$, which is less reliable than the method that uses parallaxes, especially considering that $\log g$ is a much more difficult parameter to determine with high accuracy than the parallaxes of relatively bright nearby stars.

Therefore, the derived parameters for our sample stars are consistent with previous determinations available in the literature.

**Table B1.** Average differences between spectroscopic parameters, A(Li), masses, and ages determined in this work and in previous surveys. Values in parenthesis represent the standard deviation of the differences.

| Work | LTE A(Li) (dex) | NLTE A(Li) (dex) | $T_{eff}$ (K) | log $g$ (dex) | $v_t$ (km s$^{-1}$) | [Fe/H] (dex) | Age (Gyr) | Mass (M$_\odot$) |
|---|---|---|---|---|---|---|---|---|
| Ramírez et al. (2012) | – | − 0.19 (0.31) | − 8.6 (27.4) | − 0.08 (0.05) | – | − 0.0069 (0.034) | – | – |
| Delgado Mena et al. (2014) | 0.03 (0.10) | – | 7.6 (10.1) | 0.00 (0.03) | 0.073 (0.037) | 0.010 (0.013) | 0.87 (1.97) | 0.009 (0.023) |
| Bensby & Lind (2018) | – | 0.15 (0.21) | 72.5 (33.8) | − 0.0017 (0.079) | – | 0.038 (0.022) | − 4.17 (2.83) | 0.047 (0.023) |

This paper has been typeset from a T<sub>E</sub>X/L<sup>A</sup>T<sub>E</sub>X file prepared by the author.